# Analysis of contributions of plastic deformation and intergranular corrosion to corrosion-fatigue failure of Al-Mg alloys

Oleg Belkin, Vladimir Chuvil'deev, Mikhail Chegurov, Aleksey Nokhrin[1], Anatoly Sysoev

*Lobachevsky University, 603022, Russia, Nizhny Novgorod*

chuvildeev@nifti.unn.ru

**Abstract**

The article presents the results of corrosion-fatigue tests of industrial Al-Mg alloys were conducted in air and in a 3% NaCl aqueous solution. Fatigue curves can be characterized using the Basquin equation and the plastic deformation model at the crack tip. It has been demonstrated that the primary contributions to corrosion-fatigue failure in the Al-Mg alloys at low stresses are made by the process of pitting and intergranular corrosion, and by the plastic deformation at high stresses.

**Keywords**: Al-Mg alloy; fatigue; corrosion; plastic deformation.

## 1. Introduction

Replacing steel with Al-Mg alloys, while maintaining acceptable levels of specific strength and corrosion resistance, significantly reduces vehicle weight. One of the obstacles hindering the broader application of Al-Mg alloys is their low corrosion-fatigue life (CFL). This reduces the service life of aluminum products subjected to simultaneous exposure to corrosive-aggressive environment and stresses.

The fatigue resistance of Al-Mg alloys has been extensively studied, including the fatigue failure of fine-grained alloys. However, the corrosion-fatigue fracture mechanisms in Al-Mg alloys have been insufficiently studied. The question of the relationship between the contributions of plastic deformation and the corrosive environment to the CFL of Al-Mg alloys remains particularly underexplored. This complicates the development of models for predicting the service life, as well as the formulation of recommendations for optimizing the Al-Mg alloys microstructure.

The aim of this article is to study the corrosion-fatigue failure mechanisms in Al-Mg alloys with different strength and different corrosion resistance.

## 2. Materials and methods

The objects of the study were alloys Al-2.1%Mg-0.4%Mn (Russian alloy AMg2, analog of alloy 5251), Al-5.2%Mg-0.5%Mn (Russian alloy AMg5, analog of alloys 5056 and 5083) and Al-6.1%Mg-0.45%Mn-0.25%Sc-0.10%Zr alloy (Alloy 1570).

To study the microstructure of the alloys, a Leica DM IRM optical microscope (OM) and a JEOL JSM-6490 scanning electron microscope (SEM), equipped with a BSE detector and an INCA 350 EDS microanalyzer, were used. Tensile tests of cylindrical samples were performed on a 2167 R-50 machine. The yield strength (YS), ultimate tensile strength (UTS), elongation to failure ($\delta_5$), and contraction to failure ($\psi$) were determined during testing. The methodology of stress-relaxation compression tests [1] was used to determine the macroelastic limit ($\sigma_0$) and the physical yield strength ($\sigma_y$). During the tests, the dependence of the stress-relaxation on the summary stress $\Delta\sigma_i(\sigma)$ was recorded, on which the regions of macroelastic, microplastic, and macroplastic deformation were identified [1].

---
[1] Corresponding author (nokhrin@nifti.unn.ru)

Smooth cylindrical samples were used for corrosion-fatigue tests. The rotating bending test were carried out in air and in a 3% NaCl aqueous solution. The loading frequency was 50 Hz; the loading cycle asymmetry coefficient $K = -1$. Stationary corrosion tests without load and three-point bending tests at 200 MPa were carried out in a 3% NaCl aqueous solution induces intergranular corrosion (IGC) in Al-Mg alloys. The tests were conducted at room temperature and their duration was 24 h.

## 3. Results

The average grain size of Al-2Mg, Al-5Mg and Al-6Mg-Sc-Zr alloys is ~50-100, 50-70 and 5-10 µm, respectively.

Alloys contain inclusions of two types. These are light particles ranging in size 1-10 µm (particles $A$, $Al_6Mn$-based phase) and dark particles ranging in size 3-5 µm (particles $B$, $Mg_2Al_3$-based β-phase) (Fig. 1a, c). The particles $A$ in Al-2Mg and Al-5Mg alloys contain 1.3-3.8%Mg, 1.6-3.7%Si, 3.5-10%Mn, and 9.6-19.8%Fe. The particles $A$ in Al-6Mg-Sc-Zr alloy also contain 20.9-21.5%Sc, 13.4-16.8%Zr, 2.3-3.1%Ti, 0.5%Cu. The volume fraction of $A$ particles in all alloys is ~ 3 %. The particles $B$ contain 10.1-26.8%Mg, 8.8-21.4%Si, and 0.3%Mn. The β-phase are located along the grain boundaries of the alloys. The volume fraction of β-phase increases with increasing Mg concentration from ~ 0.2 % in Al-2Mg alloy to 0.5–1 % in Al-6Mg-Sc-Zr alloy.

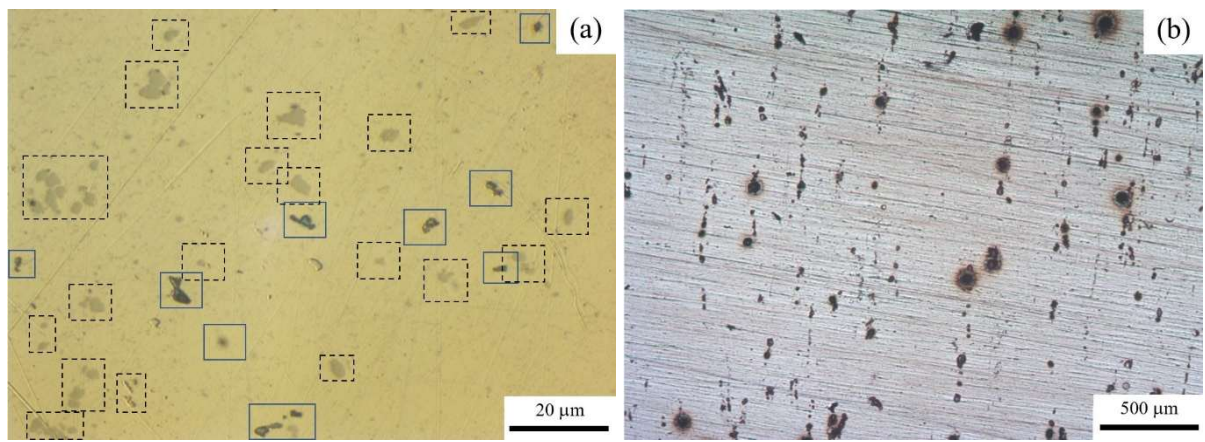

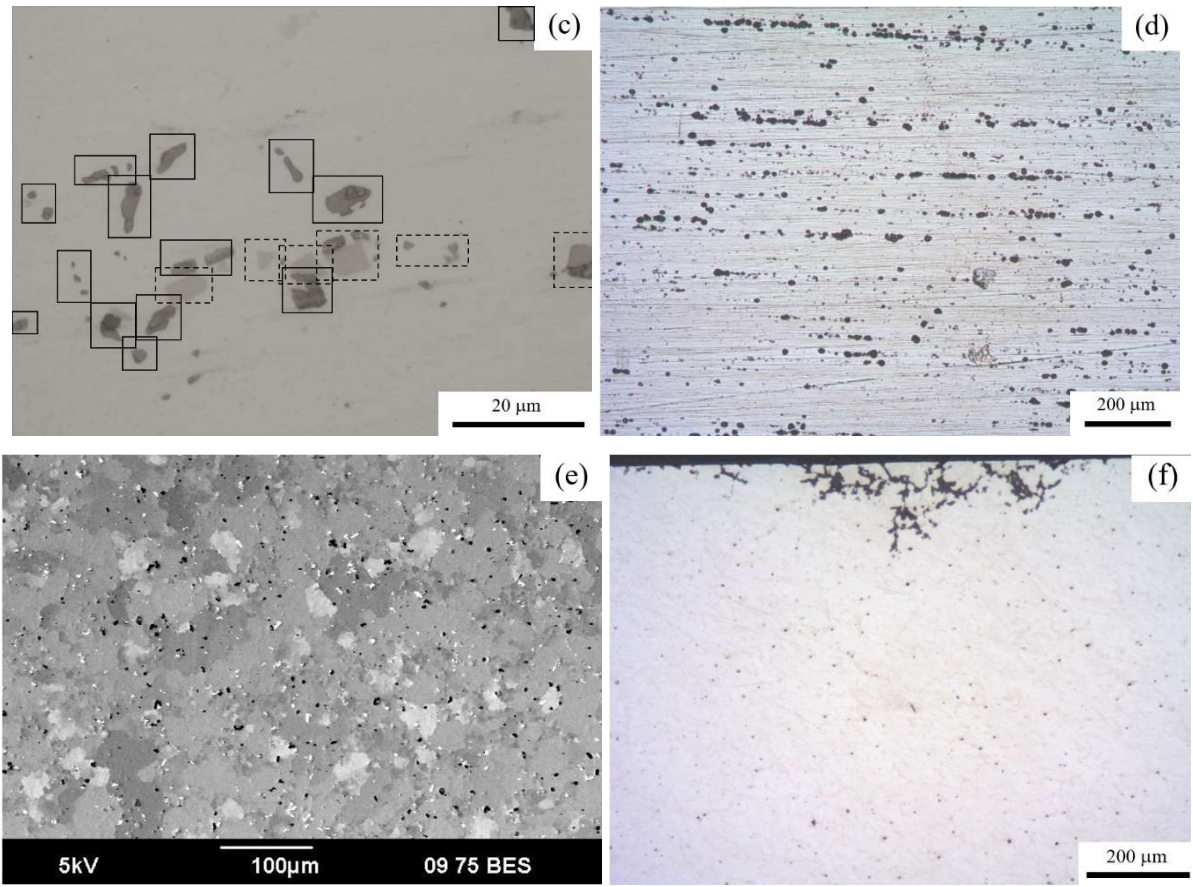

**Fig. 1** Surface of Al-5Mg (a, b, e) and Al-6Mg-Sc-Zr (c, d, f) alloy samples after polishing (a, c) and after corrosion test (b, d, e). Fig. 1e shows the alloy microstructure, and Fig. 1f – a cross-section of the grinding with IGC defects. The particles *A* are circled by dashed lines, while particles *B* are outlined by solid lines

Fig. 2a, b illustrates the tensile curves and the compression curves of the Al-Mg alloys. Table 1 shows that an increase in the YS and the UTS of the alloys is observed with increasing Mg content.

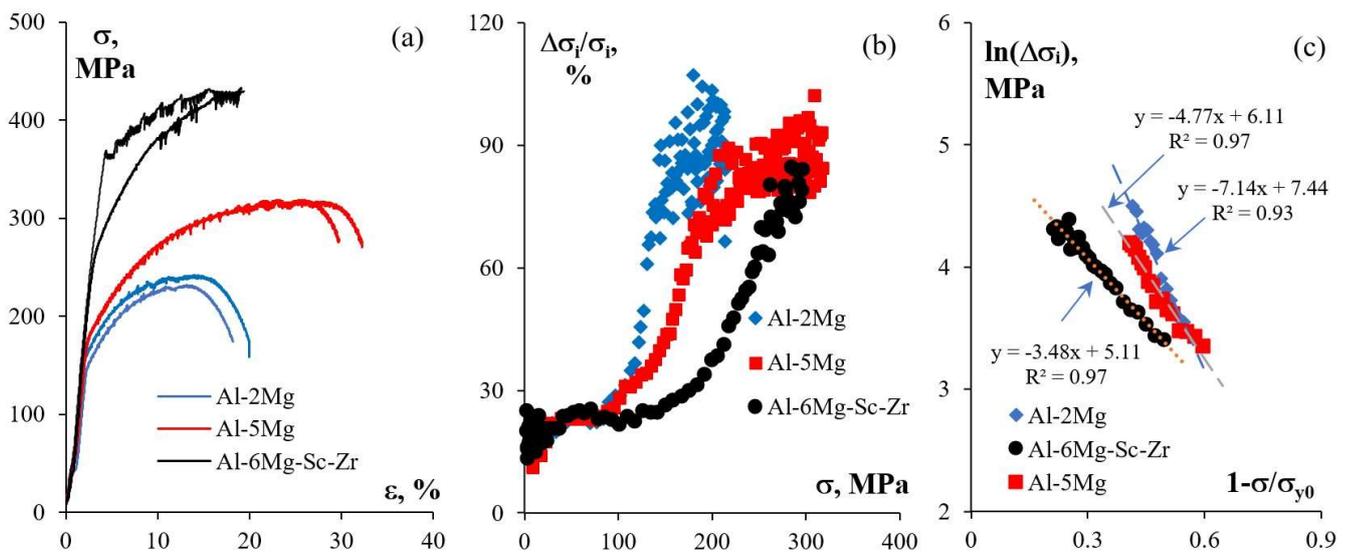

**Fig. 2** Results of tensile (a) and compression (b) tests on Al-Mg alloys. Fig. 2c - $\Delta\sigma_i(\sigma)$ curves in coordinates of $\ln(\Delta\sigma_i)$-$(1-\sigma/\sigma_{y0})$

Table 1. Results of tests of Al-Mg alloys

| Alloy | Compression tests | | Tensile tests | | | | Fatigue tests | |
|---|---|---|---|---|---|---|---|---|
| | | | | | | | Air | NaCl |
| | $\sigma_0$, MPa | $\sigma_y$, MPa | YS, MPa | UTS, MPa | $\delta_5$, % | $\psi$, % | $\sigma_{-1}$, MPa | |
| Al-2Mg | 60-65 | 150-160 | 150-155 | 230-240 | 17-19 | 45-54 | 100±15 | 60±10 |
| Al-5Mg | 70-75 | 200-210 | 175-180 | 315-320 | 26-29 | 36-40 | 120±15 | 75±10 |
| Al-6Mg-Sc-Zr | 115-120 | 280-290 | 340-365 | 430-435 | 14-19.5 | 12.5-19.5 | 155±20 | 85±10 |

Fig. 1b, d, f presents typical images of the sample surfaces after corrosion tests. The large corrosion pits and fine IGC defects are observed in the alloy samples. A large number of pits and IGC defects have been observed on the surface of Al-6Mg-Sc-Zr alloy samples, with sizes reaching up to hundreds of μm.

Fig. 3a, b shows the results of fatigue tests of the alloys. The S-N curves can be accurately interpolated using a Basquin equation: $\sigma_a = A \cdot N^{-k}$. The fatigue limits $\sigma_{-1}$ based on $10^7$ cycles, calculated using the Basquin equation, are shown in the Table 1. The increase in the strength of the Al-Mg alloys was accompanied by an increase in the $\sigma_{-1}$.

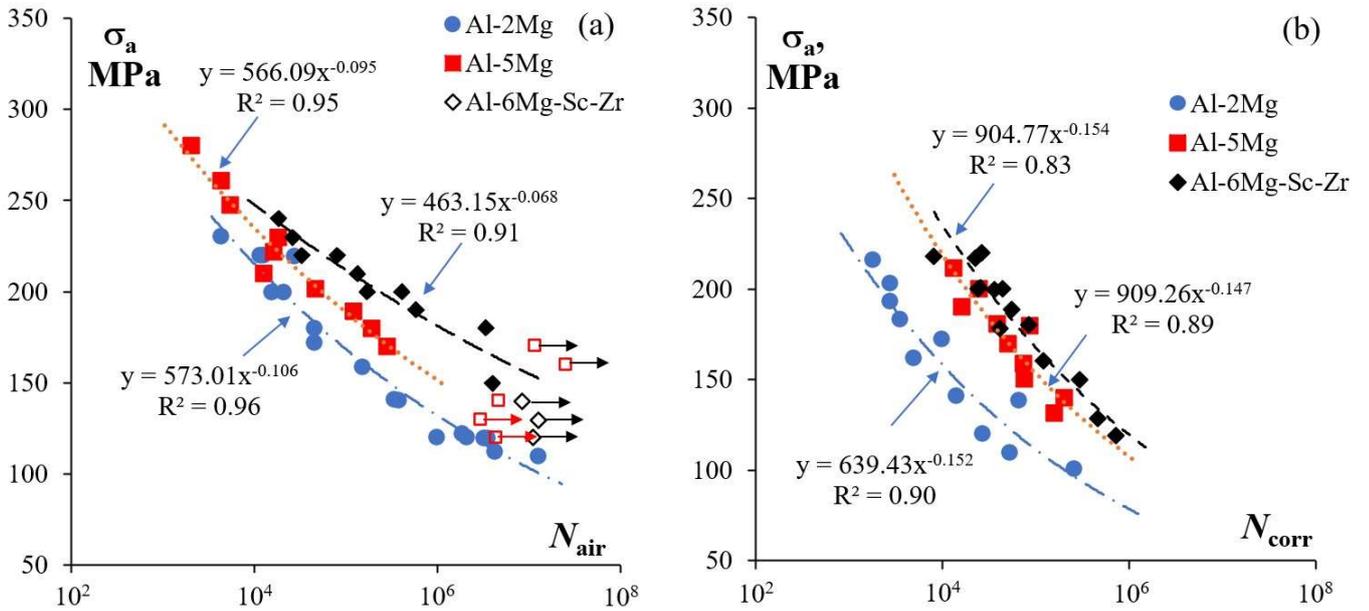

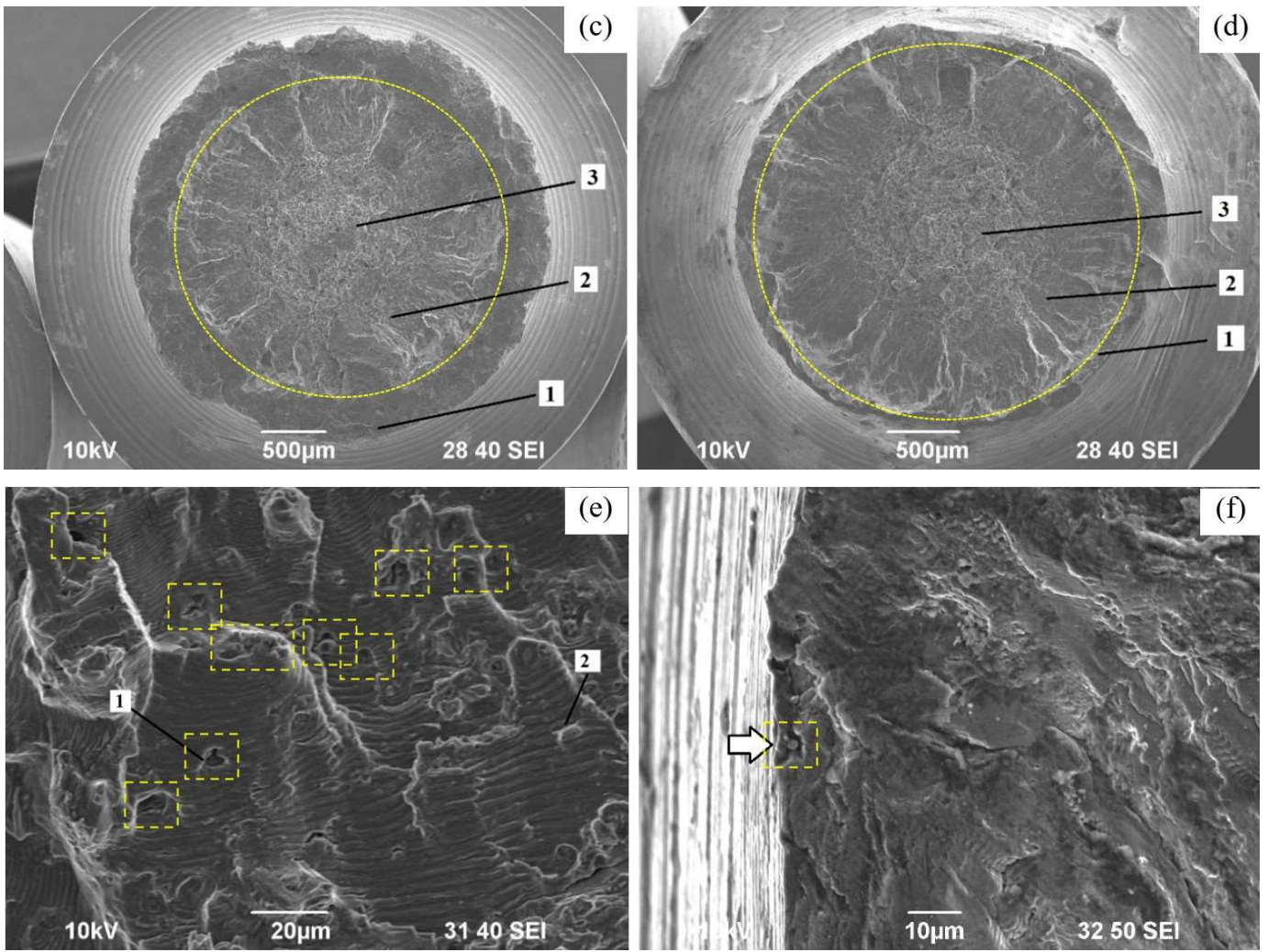

**Fig. 3** Results of fatigue test: (a, b) S-N fatigue (a) and corrosion-fatigue (b) curves for Al alloys; fractures after corrosion-fatigue tests at 160 MPa (c) and 215 MPa (d); (e, f) inclusions in the zone of accelerated crack growth (e) and on the crack origin zone (f) after fatigue (e) and corrosion-fatigue tests (f). SEM

Note also that alloy Al-6Mg-Sc-Zr usually has higher $\sigma_{-1}$. In its initial state, the value of $\sigma_{-1}$ is close to 200 MPa, and after strain hardening treatment, it exceeds 250-300 MPa [3]. A similar observation applies to alloy 5056 [4]. The primary reason for the observed contradiction is the presence of inclusions in the alloys, which can serve as nucleation areas of microcracks.

The corrosion-fatigue curves are positioned lower than the S-N curves observed in air. The CFL of the Al-Mg alloys is lower than in air tests (Fig. 3a, b). It is noteworthy that the difference in the $\sigma_{-1}$ obtained from corrosion-fatigue tests for the alloys Al-2Mg and Al-6Mg-Sc-Zr is minimal, only slightly exceeds the uncertainty in the determination of $\sigma_{-1}$. The $\sigma_{-1}$ observed in corrosion-fatigue tests is 1.3-1.8 times lower than that in air tests (Table 1).

SEM-analysis reveals that the fatigue fracture surface of Al-Mg alloys exhibits three distinct zones: (1) a zone of crack origin and of slow crack growth; (2) a accelerated crack growth zone; (3) the final fracture zone. As the $\sigma_a$ increases, there is an increase in the number of crack nucleation centers.

The samples after corrosion-fatigue testing indicates that the same three standard zones are observed on the fracture surface as on the surface of the samples after testing in air. The crack nucleation zone on the fracture of the samples tested in the corrosive environment at low $\sigma_a$ is uniformly distributed over the entire diameter of the sample (Fig. 3c). In contrast, when tested in air, this zone is localized to specific areas of the fracture. As the $\sigma_a$ increases, the area of microcrack initiation decreases (Fig. 3d). The areas of fracture of inclusions appear on the fracture surface of the samples after corrosion-fatigue testing (Fig. 3f).

## 4. Discussion

The model of dislocation slidings [5] can be used to describe microplastic deformation in coarse-grained materials. According to [1], the slope angle ($K_1$) of the $\Delta\sigma_i(\sigma)$ curves in the $\ln(\sigma_i)$-$(1-\sigma/\sigma_{y0})$ coordinates corresponds to the inverse of the microplastic deformation activation energy ($\Delta F_1$): $K_1 = kT/\Delta F_1$, where $\sigma_{y0}$ is the flow stress at 0 K, $k$ is the Boltzmann constant, and $T$ is the temperature. The primary reason for the increased values of the $\Delta F_1/kT$ in alloy Al-6Mg-Sc-Zr (Fig. 2c) is the presence of $Al_3(Sc,Zr)$ particles and the high concentration of Mg atoms in the crystal lattice.

According to the models [6, 7], the strain rate ($\dot{\varepsilon}_v$) determines the time to microcrack nucleation and the microcrack growth rate. In this case, the slope angle of the S-N curve corresponds to the parameter $A$ in the Basquin equation, which is proportional to the microplastic deformation activation energy at the crack tip ($\Delta F_2/kT$) [6].

It should be noted that the absence of a noticeable change in the $A$ in the case of fatigue tests is an unexpected finding. As previously demonstrated, alloy Al-6Mg-Sc-Zr exhibits a higher $\Delta F_1/kT$ and is characterized by higher values of the $\sigma_{-1}$ compared to alloys Al-2Mg and Al-5Mg. The primary reason for the observed contradiction is the effect of accelerated crack growth in fine-grained materials [8].

The decrease in the $\Delta F_2/kT$ observed during tests in a corrosive environment can be attributed to the Rehbinder effect. The second factor contributing to a decrease in the $\Delta F_2/kT$ is the presence of a corrosion defects on the surface. Such defects act as stress concentrators, leading to a localized increase in the $\dot{\varepsilon}_v$. This assumption is supported by the relationship between the difference in the life in air ($N_{air}$) and in a corrosive environment ($N_{corr}$) and the stress amplitude. Fig. 3a, b shows that decreasing the $\sigma_a$ leads to an increase in the $\Delta N = N_{air} - N_{corr}$. A decrease in the $\sigma_a$ results in an increased exposition time of the sample in the corrosive environment, which consequently leads to an increase in the number and size of corrosion defects. This leads to an increase in the number of the crack nucleation centers, which are uniformly distributed over the sample diameter (Fig. 3c). At the same time, an increase in the number and size of defects leads to a local increase in the $\dot{\varepsilon}_v$. At a given $\sigma_a$, this is equivalent to a decrease in the $\Delta F/kT$.

## 5. Conclusions

- An increase in the strength of the Al-Mg alloys is accompanied by an increase in the fatigue limit $\sigma_{-1}$ and a slight decrease in the parameter $A$ in the Basquin equation, which corresponds to the plastic deformation activation energy at the crack tip: $A \sim (\Delta F/kT)^{-1}$. The small scale of change in $\Delta F/kT$ is due to the effect of crack growth acceleration in fine-grained materials.
- An increase in the strength of the Al-Mg alloys does not enhance the $\sigma_{-1}$ value in corrosion-fatigue tests. The reasons for the decrease in $\sigma_{-1}$ and $\Delta F/kT$ are as follows the Rehbinder effect and the formation of corrosion defects on the inclusions and on the β-phase particles.
- At low stress amplitudes, the primary contributions to corrosion-fatigue life in the Al-Mg alloys are pitting and IGC, and at high stresses – plastic deformation at the crack tip.